\documentclass[preprint,superscriptaddress, amsmath,amssymb, aps,floatfix,]{revtex4-2}

\usepackage{graphicx}
\usepackage{dcolumn}
\usepackage{bm}
\usepackage{caption}
\usepackage{subcaption}
\usepackage{hyperref}

\usepackage{siunitx}
\usepackage{verbatim}

\newtheorem{proof}{Proof}

\begin{document}


\title{Mesoscopic analytical approach in a three state opinion model with continuous internal variable}

\author{Luc\'ia Pedraza}
\email{lpedraza@df.uba.ar}
\affiliation{Departamento de F\'isica, Facultad de Ciencias Exactas y Naturales, Universidad de Buenos Aires. Av.Cantilo s/n, Pabell\'on 1, Ciudad Universitaria, 1428, Buenos Aires, Argentina.}
\affiliation{Instituto de F\'isica de Buenos Aires (IFIBA), CONICET. Av.Cantilo s/n, Pabell\'on 1, Ciudad Universitaria, 1428, Buenos Aires, Argentina.}

\author{Juan Pablo Pinasco}
\affiliation{Departamento de Matem\'atica, Facultad de Ciencias Exactas y Naturales, Universidad de Buenos Aires and IMAS UBA-CONICET, Av. Cantilo s/n, Pabell\'on 1, Ciudad Universitaria, 1428, Buenos Aires, Argentina.}
\author{Viktoriya Semeshenko}
\affiliation{Universidad de Buenos Aires. Facultad de Ciencias Econ\'omicas. Buenos Aires, Argentina. Instituto Interdisciplinario de Econom\'ia Pol\'itica de Buenos Aires. Buenos Aires, Argentina}

\author{Pablo Balenzuela}
\affiliation{Departamento de F\'isica, Facultad de Ciencias Exactas y Naturales, Universidad de Buenos Aires. Av.Cantilo s/n, Pabell\'on 1, Ciudad Universitaria, 1428, Buenos Aires, Argentina.}
\affiliation{Instituto de F\'isica de Buenos Aires (IFIBA), CONICET. Av.Cantilo s/n, Pabell\'on 1, Ciudad Universitaria, 1428, Buenos Aires, Argentina.}

\begin{abstract}
Analytical approaches in models of opinion formation have been extensively studied  either for an opinion represented as a discrete or a continuous variable. In this paper, we analyze a model which combines both approaches. The state of an agent is represented with an internal continuous variable (the leaning or propensity), that leads to a discrete public opinion: pro, against or neutral. This model can be described by a set of master equations which are a nonlinear coupled system of first order differential equations of hyperbolic type including non-local terms and non-local boundary conditions, which can't be solved analytically. We developed an approximation to tackle this difficulty by deriving a set of master equations for the dynamics of the average leaning of agents with the same opinion, under the hypothesis of a time scale separation in the dynamics of the variables. We show that this simplified model  accurately predicts the expected transition between a neutral consensus and a bi-polarized state, and also gives an excellent approximation for the dynamics of the average leaning of agents with the same opinion, even when the time separation scale hypothesis is not completely fulfilled.
\end{abstract}

\maketitle

\section{Introduction}
\label{sec:intro}

Opinion formation processes play an important role in societies. Many societies seem to shift towards more polarization and volatility in opinions, for instance, in issues about immigration \cite{Ianellietal2021}, climate denial change \cite{Aaron2022}, or, more recently, the responses to policy measures to Covid-19 pandemics \cite{Milligan2020, Roberts2020}. The underlying dynamics behind these opinion formation processes are highly complex due to the interdependence of interconnected individuals influencing each other in different ways \cite{flache2017, kozitsin2022formal}. 

Understanding how the opinions are formed is not an easy task as the dynamics of how one changes his opinion based on his interactions with others is unclear and not trivial at all. It is necessary to understand opinion dynamics through a complex micro-macro social processes which occur at different levels: micro-level like individuals, meso-level like network structures, and macro-level outcomes, like consensus, opinion-clustering or polarization. 

This, in turn motivated researchers from different disciplines to investigate and explore the opinion dynamics, and we can find many contributions from sociology \cite{Schelling,Granovetter}, psychology \cite{Vickers,Smith}, data analysis \cite{Cinelli}, and statistical physics \cite{Abelson, DeGroot, Weisbuch2, Burnstein,Sunstein,Sampedro,Dandekar,Krueger, Jager}. 

Over the past decades, great advances have been made, and a large variety of opinion formation models have been proposed. The models may differ in the way the opinions are represented. Opinions could be modeled in a discrete space as a set of options (for example, a pro/against issues), or in a continuous space, as a real number in a finite interval which represents the orientation between two given extremes.

Some discrete models are based on analogies from condensed matter physics, with opinions treated as discrete states, resembling spin states in solids. The changes of opinions in such models are typically attributed to pairwise interactions between agents or the influence of groups of agents or external media on a single agent. The external influence is introduced through parameters corresponding to an external magnetic field, and the volatility in individual opinions is associated with ``social temperature". 
These models vary considerably in the ways the interactions between agents are described. 
Among the most popular ones, we can mention the voter model \cite{ben1996coarsening, Clifford,Holley,Cox,Liggett,Sire}, and the Sznajd model \cite{sznajd2000opinion}. 

The common feature of these approaches is in the form in which opinion change is modeled, i.e. an agent's opinion change depends on the combination of his current opinion and the opinions of his neighbors and external influences. While the models vary in details, a frequent assumption is that an agent may change his opinion as long as there are one or more disagreeing agents in a neighborhood, after a single interaction with another agent, or when the local majority of agents favors a different opinion. Using the analogy with physics, these models focused on conditions in which society would achieve consensus. 

Agent model simulations are commonly the tools to explore these types of models. However, analytical developments allow us to calculate a master equation that predicts the density of opinions as a function of time. The study of this equation can be an alternative way to study the different final states depending on the parameters as well as the transitions. These equations predict dynamics well compared to simulations for models with simple interactions. 

Classical continuous models were grounded in different kinds of microscopic interactions. These models usually consider a discrete time,  at each time agents change their opinions influenced by others \cite{DeGroot}. A continuous representation allows introducing a bounded confidence limit to represent that agents only interact if their similarity is lower than a given threshold \cite{deffuant2000mixing, weisbuch2004bounded, lorenz2007continuous}, the Hegselmann-Krause model \cite{hegselmann2002opinion}. 
The opinions of two interacting agents could get closer to each other only if the initial difference is acceptable small, below the tolerance threshold. Depending on this tolerance threshold, this could lead to consensus (if the tolerance is high) or a stable coexistence of two or more groups, within which the opinion would converge to different values. 
These models were extensively studied in the last years \cite{amblarddeffuant2004, deffuant2006}. They could be analyzed either with agent-based dynamics for a finite
number of $N$ interacting agents, or with partial differential equations governing the evolution of a density function that represents the distribution of agents in the opinion space. These partial differential equations can be obtained as approximations of Boltzmann type equations modeling the agents' interactions, as was shown in \cite{Bellomo, Pareschi, Toscani}.

Nevertheless, to explain a complex dynamic as opinion formation and interactions between persons, more complex models are needed. The development of the master equations could become difficult and often impossible for increasing complexity of the model, and approximations may result useful.

In \cite{BPS,BSNB}, authors combine both approaches and represent the state of an agent with an internal continuous variable, that leads to a discrete public opinion among three options: pro, against, or neutral. The behaviour of the agents depends on this discrete opinion. The authors prove that the neutral state is fundamental to reach polarization. The introduction of neutral or undecided agents has been also studied in \cite{deLaLama, Couzin, Sobkowicz, Vazquez, Svenkeson, Singh,Marvel} while the study of agents divided in three groups or parties has been studied in \cite{Galam1, Galam2, Galam3, Galam4}.
Also, in \cite{BPS,BSNB}, the authors were able to deduce the master equations corresponding to the dynamics of the system. These master equations are a nonlinear coupled system of first order hyperbolic differential equations including non-local terms and non-local boundary conditions. The master equations are of special interest for their nontrivial properties and difficulties in being solved analytically. 

In this work, we perform an approximation to resolve analytically the model proposed in \cite{BPS}. 
We develop a set of equations by doing a simplification of the model focusing on the dynamics of the average leaning of agents with the same opinion. We consider that for a range of parameters the dynamics occur in two different time scales. There is a fast dynamic in which each set of agents with the same opinion tends to have the same leaning and a slow dynamic in which the mean of the leaning of each group of agents with the same opinion changes due to interactions with agents with different opinions. 

We deduce these equations explicitly and verify that they are an excellent approximation of the dynamics of the model, providing a simple and accurate description in different scenarios.

This work shows the strength of this approach for the analytical treatment of opinion models with mixed variables (continuous underlying and discrete emerging).

The article is organized as follows: the model and the interaction dynamics are presented in section \ref{sec:model}. The approximation used for analytical treatment is in section \ref{sec:approx}. The results are presented and discussed in section \ref{sec:results}. We conclude in section \ref{sec:conclu} and provide a theoretical development of master equations in Appendix \ref{appendix:equations}.

\section{Model}
\label{sec:model}

\begin{figure}[htp]
    \centering
    \includegraphics[width =0.6\columnwidth]{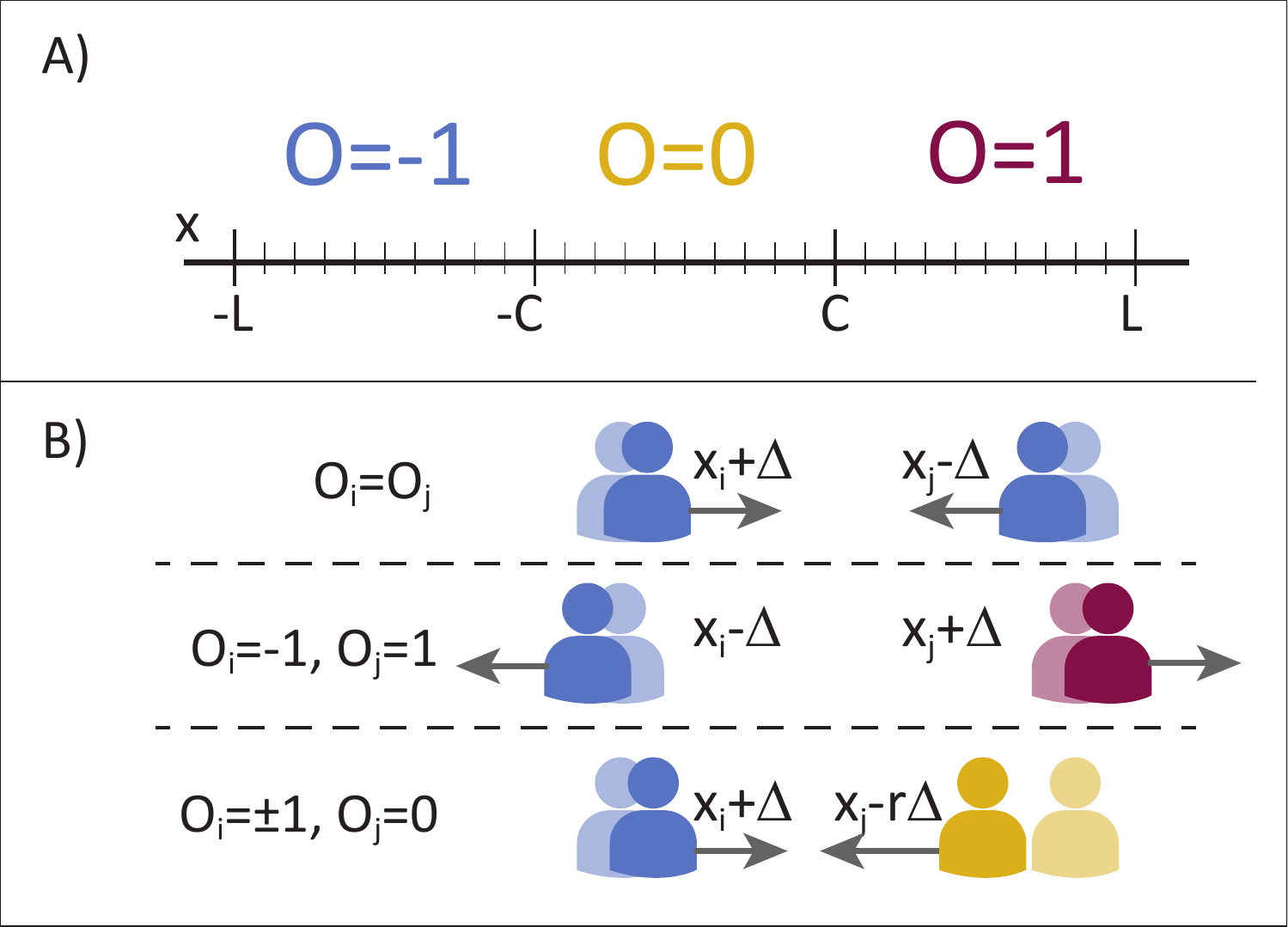}
    \caption{ Panel A: The definition of each agent's state. Agents with  leanings in the intervals $[-L,-C]$,$[-C,C]$ and $[C,L]$ have  opinion $O=-1$, $O=0$ and $O=1$ respectively. Panel B: Pair-wise interaction rules. Agents with the same opinions get closer after an interaction, while agents with extreme and different opinion distance themselves. Agents with neutral opinions, which participate in the interaction,  get attracted but asymmetrically.
    }
    \label{Fig1}
\end{figure}

The model considers $N$ interacting agents. The states of agents are determined by two interdependent variables:  the  \textit{leaning}, represented by a continuous variable $x \in [-L,L]$, $L \in \mathbb{R}$, and the \textit{opinion} $O$, which can be either positive, negative or neutral. Both are interrelated by \textit{thresholds} $\pm C$ (see Fig.\ref{Fig1}(A)). If an agent's leaning $x$ is in $[-C,C]$, the agent will have a neutral opinion $O=0$. If $x<-C$ he gets a negative opinion $O=-1$, and otherwise, if $x>C$, the agent gets a positive opinion $O=1$. We say that agents with the same opinion belong to the same \textit{community}. So far, there are three communities $\{-1,0,1\}$ in the model.

The dynamic is given by pairwise interactions. At each time step  $\tau$, two agents are randomly selected to interact, and the leanings of both are modified by a small $\Delta$, following the rules (see Fig.\ref{Fig1}(B)): 

\begin{itemize}
\item If agents share the same opinions, $O_i(t)=O_j(t)$, and $x_i(\tau)>x_j(t)$, then they approach each other; like-minded agents influence each other in a way that their leanings become more closer:
\begin{align*}
x_i(t+\tau)&=x_j(t)-\Delta ,\\
x_j(t+\tau)&=x_i(t)+\Delta .
\end{align*}

\item If agents hold different extreme opinions (without loss of generality $O_i(t)=+1$, and $O_j(t)=-1$) then they get repulsed; agents move further away from each other:

\begin{align*}
x_i(t+\tau)&=x_j(t)+\Delta ,\\
x_j(t+\tau)&=x_i(t)-\Delta .
\end{align*}

\item If one of the agents holds either of extreme opinions and the other one is neutral, $O_i(t)=\pm 1$ and $O_j(t)= 0$, then agents approach each other asymmetrically:
\begin{align*}
x_i(t+\tau)&=x_j(t)\mp\Delta ,\\
x_j(t+\tau)&=x_i(t)\pm r\Delta, 
\end{align*}
 where $r>1$ accounts for the mentioned asymmetry.
\end{itemize}

In turn, if after an interaction, agents' leanings cross the threshold $C$, the opinions will also change, and the number of agents in each community will be modified. 
The macroscopic states of the system can be described in terms of the fraction of agents in each community: $n_+(t)$, $n_-(t)$ y $n_0(t)$ (for positive, negative, or neutral respectively). 

We set the initial proportion of agents in each community as a function of the parameters $n^i_0$ (the initial proportion of neutral agents) and $d$ (the difference between the proportions of positive and negative communities):

\begin{equation}
\begin{split}
n_0(0)&=n^i_0,\\
n_+(0)&=(1-n^i_0)/2+d,\\
n_-(0)&=(1-n^i_0)/2-d.
\label{condicion_inicial}
\end{split}    
\end{equation}

At equilibrium ($t_f$), there are three possible final states:

\begin{itemize}
    \item {\underline {Bi-polarized state}}. There are two extreme communities: $0<n_+(t_f)<1$ and $0<n_-(t_f)<1$, and no neutral agents, $n_0(t_f)=0$.   
    \item {\underline {Neutral consensus}}. The entire community is moderate: $n_0(t_f)=1$.
    \item {\underline {Extreme consensus}}. The entire community is extreme, either negative or positive. $n_+(t_f)=1$ or $n_-(t_f)=1$.
\end{itemize}

Given the description of the system, its dynamic rules and equilibrium states, we can now look at the equations which rule systems dynamics. 
In the limit of $\Delta \rightarrow 0$ and $N \rightarrow \infty$, an analytical approach can be derived in terms  of the density of the agent's leaning: $f(x,t):[-L,L]\times [0,t_f]\rightarrow [0,1]$, which can be used to calculate the fraction of agents in each community:
 
\begin{equation}
\begin{split}
n_-(t)&=\int_{-L}^{-C} f(x,t) dx,\\
n_0(t)&=\int_{-C}^C f(x,t) dx,\\
n_+(t)&=\int_{C}^L f(x,t) dx.
\end{split}
\end{equation}

Given these densities, a master equation can be  derived (as in  \cite{BPS}): 

\begin{equation}
\frac{\tau N}{2\Delta} \frac{d}{dt}f(x,t)
= \left\{ \begin{array}{lcc}
&\frac{d}{dx}\Big[   f(x,t) \Big( \int_{-L}^x f(y,t)dy - \int_x^{-C} f(y,t)dy\Big)  \Big] \\
&+\frac{d}{dx} f(x,t) (n_0(t) - n_+(t))  & for \quad  x \in[-L,-C], \\
 &\frac{d}{dx}\Big[   f(x,t) \Big( \int_{-C}^x f(y,t)dy - \int_x^C f(y,t)dy\Big)  \Big] \\ 
 &+\frac{d}{dx}  f(x,t) ( n_+(t) - n_-(t))  & for \quad  x \in[-C,C],\\
 &\frac{d}{dx}\Big[  f(x,t) \Big( \int_C^x f(y,t)dy - \int_x^L f(y,t)dy\Big)  \Big] \\
&+r\frac{d}{dx}   f(x,t) ( n_-(t) - n_0(t) )  & for \quad x \in[C,L].
\end{array}
\right.
\label{ecuaciones_generales}
\end{equation}

For the sake of simplicity, we can set the time scale of the rate of interactions  as $\tau N = 2\Delta$ (we omitted the boundary conditions).  

\begin{figure}[htp]
    \centering
    \includegraphics[width =\columnwidth]{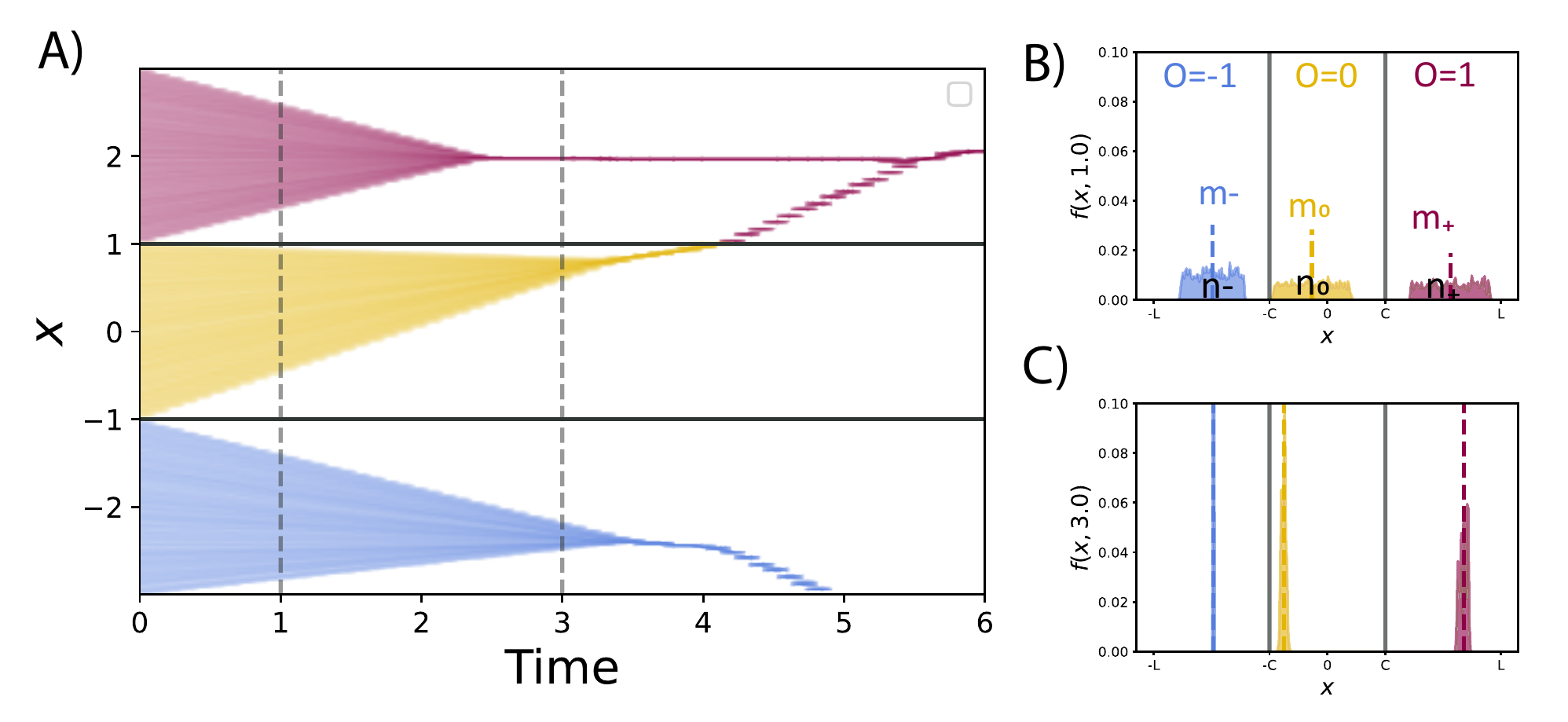}
    \caption{ Panel A: Leaning dynamics for the system with $N=10000$, $L=3$, $C=1$, initial fractions of agents in each communities $n_-(0)=0.29$, $n_0(0)=0.3$, $n_+(0)=0.41$, uniformly distributed in each community. Panel B and C: Agent's leaning distribution for the same system realization at two different times steps, $t=1$ (panel B) and $t=3$ (panel C).
    }
    \label{Fig2}
\end{figure}

Dealing with these types of equations is extremely hard and, in some cases, it is impossible to get the exact solution. 
However, if we observe the dynamics of agent's leaning as is sketched in Fig.\ref{Fig2}(A), we can see two clear time scales: a fast one where agents with the same opinion converge to their mean average leaning, as was described in \cite{PSB}, and a slow one, where agent with the same opinion has roughly the same leaning and interact as a whole.

This behavior is the cornerstone of our approach because it means that the dynamics of all agents can be approximated by the average leaning of agents with the same opinion, reducing drastically the dimension of the problem.

Here, we have to tackle two questions: 
\begin{enumerate}
    \item  Under which conditions the dynamics of all agents with the same opinion can be approximated by the one of their average leaning? 
    \item Is it possible to deduce dynamical equations for  average leaning of agents with same opinion departing from Eq. \ref{ecuaciones_generales}?
\end{enumerate}
For the first one, we provide an intuitive explanation as to why this is a good approach (a formal proof can be found in Appendix \ref{appendix:time}). 
An agent can interact with any of three communities, each in a different way: agents' leanings will get closer if they are from the same community, and will move closer or further away if they come from  different communities depending on which one. 
For instance, for an agent with a negative opinion, interaction with a neutral agent moves him closer to the neutral opinion, while interaction with an agent with a  positive opinion reinforces his negative opinion. 
If the number of agents in each community is the same, considering that interacting pairs are randomly selected at each time, we could assume that these two effects from interactions cancel each other out. Therefore, the only visible effect will be the one produced by the interaction between agents of the same community, which, due to the previous result, produces a concentration. 
If instead, the initial fractions of agents in the communities are similar but not equal, it is intuitive to assume that the effects of interaction between agents of the different communities will be mostly balanced.

For the second question we provide, in the next section, the explicit master equations for the dynamics of the average leaning of agents with the same opinion departing from Eq. \ref{ecuaciones_generales}.

\section{Master equations}
\label{sec:approx}
Let us define the average leaning of agents of each community as 
\begin{equation}
\begin{split}
m_-(t)&=\frac{1}{n_-(t)}\int_{-L}^{-C} xf(x,t) dx,\\
m_0(t)&=\frac{1}{n_0(t)}\int_{-C}^C xf(x,t) dx,\\
m_+(t)&=\frac{1}{n_+(t)}\int_{C}^L xf(x,t) dx.
\end{split}
\end{equation}

Then, a community can change its average leaning by two processes:

\begin{itemize}
\item An agent's leaning changes, but its opinion  remains unchanged. 

\item An agent's opinion changes. In this case the average leaning of both communities changes. 
\end{itemize}

Typically, the first types of changes dominate the dynamics, as we have observed in Fig.\ref{Fig2}(A). The exception occurs in a limited period of time where all agents of one community cross the threshold $C$, and all change their opinions together. At this moment, one of the communities disappears and is absorbed by another one.

The dynamics for average leaning of agents with the same opinion can be deduced from Eq. \ref{ecuaciones_generales}, as can be seen in Appendix \ref{appendix:equations}. 
Surprisingly, those lead to very simple equations: 

\begin{equation}
\begin{split}
 \frac {dm_+(t)}{dt} =&  n_-(t) - n_0(t),\\
 \frac {dm_-(t)}{dt} = & n_0(t)-n_+(t),\\
 \frac {dm_0(t)}{dt} = &  r(n_+(t)-n_-(t)).
\label{ecuacion_diferencial}
\end{split}
\end{equation}

We do not have a closed expression, since the fraction of agents in each community $ n_j(t) $ for $j \in \{+,-,0\}$ depends on time. However, if we can assume that they remain unchanged (as long as agents do not modify their opinions), we can take  $n_j(t)=n_j(0)=n_j$ for times before the first change of opinion takes place and  Eqs. \ref{ecuacion_diferencial} become linear:

\begin{equation}
\begin{split}
m_+(t)&=m_+(0)+(n_{0}-n_{-})t,\\
m_-(t)&=m_-(0)+(n_{+}-n_{0})t,\\
m_0(t)&=m_0(0)+r(n_{-}-n_{+})t .
\label{ecuacion_lineal_1}
\end{split}
\end{equation}

The dynamic of the average leaning in each community is very simple as long as agents holding the same opinion change it all simultaneously. 
Of course, this can serve as a good approximation if initially all agents with the same opinion have also the same leanings. 
But how good is this approximation when initial leanings are uniformly distributed in the three intervals? In Appendix \ref{appendix:equations}, we show that it is still a very good one as long as like-minded agents converge to their average leaning before crossing the threshold and changing opinions. 
So, the relevant times are those where the average leaning of each community reaches the threshold limits, $C$ and $-C$, and produces changes in opinions. 
Let $t_+$ and $t_-$ be the times when the average leaning of the two extremes communities brings to a neutral opinion, and $t_0^{+}$ and $t_0^{-}$ the times for the neutral community to reach the positive and negative limit, respectively. Since we assume communities move concentrated, at that time all the agents will change their opinion together. 

\begin{equation}
\begin{split}
t_+&=\frac{C-m_+(0)}{n_{0}-n_{-}},\\
t_-&=\frac{-C-m_-(0)}{n_{+}-n_{0}},\\
t_0^{+}&=\frac{C-m_0(0)}{r(n_{-}-n_{+})},\\
t_0^{-}&=\frac{-C-m_0(0)}{r(n_{-}-n_{+})}.
\label{tiempo}
\end{split}
\end{equation}

\begin{figure}[htp]
    \centering
    \includegraphics[width =0.8\columnwidth]{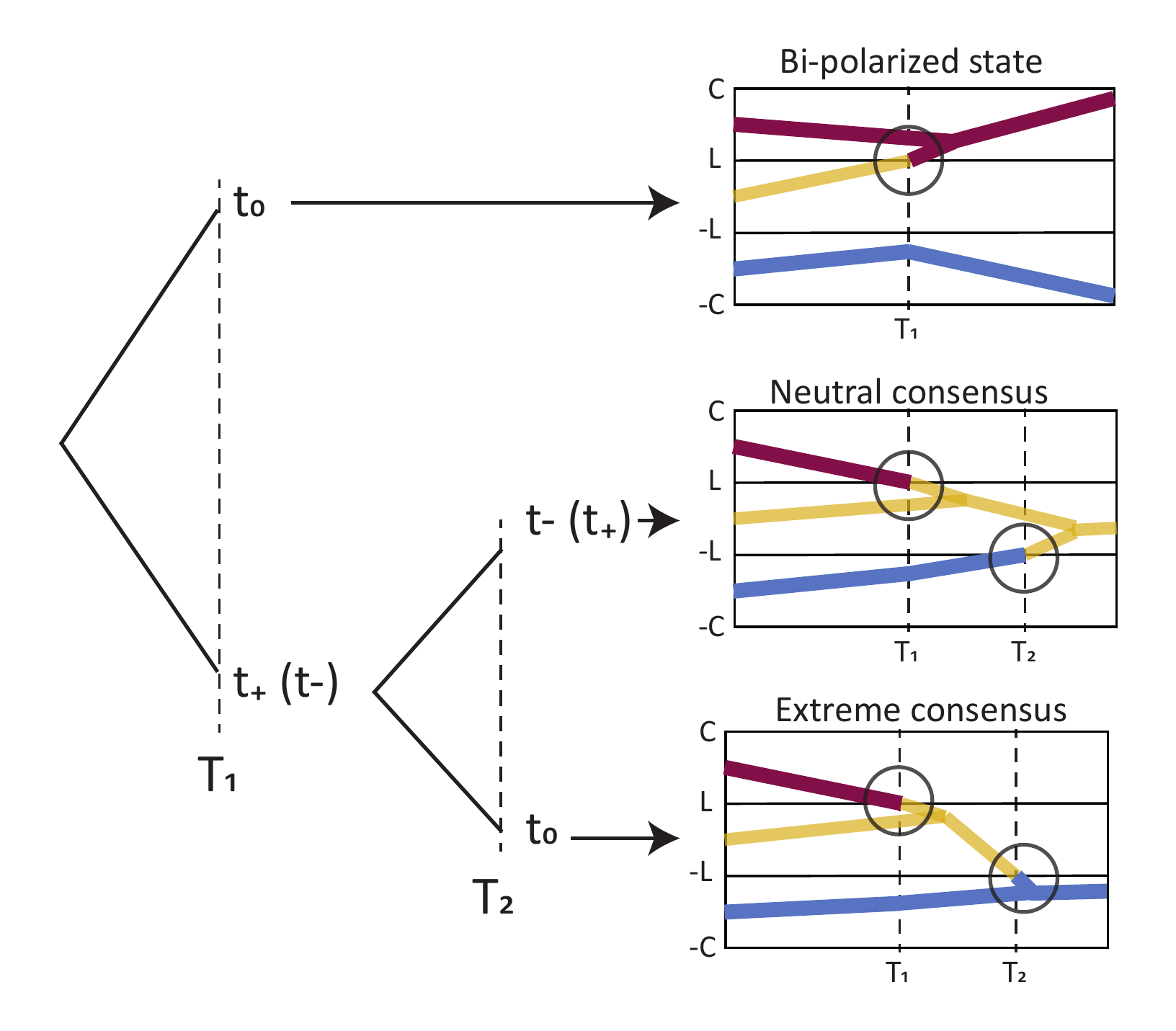}
    \caption{Scheme of the possible dynamics of the model. If $T_1=t_0$, the first community reaching a threshold is the neutral one. Therefore the final state is bi-polarized. If $T_1=t_+$ (equivalent $T_1=t_-$), an extreme community changes to the neutral one. Then the two remaining communities continue attracting and a consensus is reached. This consensus could be neutral or extreme depending on $T_2$.
    }
    \label{Fig3}
\end{figure}

We should take into account here that the only relevant solutions correspond to positive times. If any of the communities change their opinion, the dynamical equations will change. 
Let  $T_{1}=min\{t_+,t_-,t_0^{+},t_0^{-}\}$ be the time at which the first change of opinion takes place. Eqs.  \ref{ecuacion_lineal_1} will be valid until time $T_{1}$. At that moment, the entire community that reaches the threshold changes its agents' opinions and becomes another community. The possible outcomes are all sketched in Fig.\ref{Fig3}.

Fig.\ref{Fig3} shows that the simplified equations, based on the dynamics of average leanings (Eqs. \ref{ecuacion_lineal_1}), produce also the three collective states presented in the original model \cite{BPS}: Bi-polarization, neutral and extreme consensus. 
In particular, this last state (extreme consensus) only appears when the asymmetry factor in the interactions between neutral and extreme agents, $r$, is too large and initial conditions are too biased towards extreme states, that we do not take into account in this work.

For a fixed $d$ and $r$, there is a transition on $n_0$ between the bi-polarized and the consensus states. We can calculate $n_0^c$, the critical value when $t_+=t^+_0$ as: 
\begin{equation}
n^c_0=\frac{2}{3}(2r-1) d+\frac{1}{3}.
\label{Sc0_transition}
\end{equation}

For lowers values of $n_0$, $T_1=t_0^+$ and the final state will be bi-polarized. 
For higher values, $T_1=t_+$ and all agents will belong to the neutral community.

\section{Results}
\label{sec:results}

Here we compare analytical predictions of our simplified model with numerical simulations. We consider $L=3$ and the threshold $C=1$ for determining the communities. 
The initial fractions for each community are set up uniformly in the intervals $[-3,-1]$, $[-1,1]$ and $[1,3]$ respectively. 
We set the parameter $r=2$, which accounts for the asymmetric update of leanings. 

\begin{figure}[htp]
    \centering
    \includegraphics[width =\columnwidth]{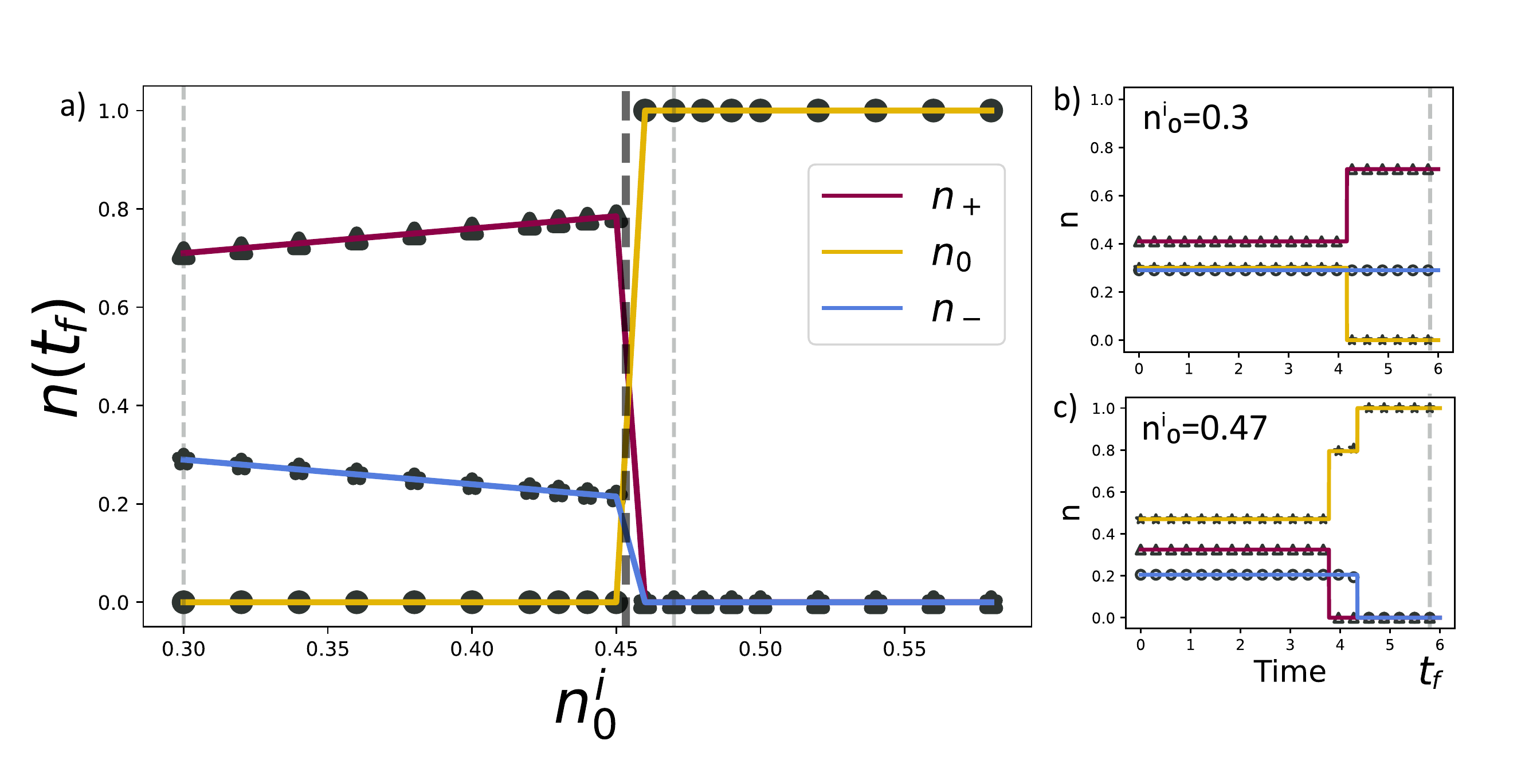}
    \caption{ Simulations (symbols) and analytical predictions (lines) for positive (violet), neutral (yellow), and negative (blue) fractions of agents in each community. Parameters used: $L=3$, $C=1$, $r=2$, $N=10000$ agents, $d=0.06$. Panel a) shows the proportion of agents at the final state. Panels b) and c) show the temporal evolution for $n^i_0=0.3$ (b) and $n^i_0=0.47$ (c). 
    }
    \label{fig4}
\end{figure}

In Fig.\ref{fig4}(a), we compute the final fraction of agents in each community as a function of the initial proportion of neutral agents, $n^i_0$ for $d=0.06$. We observe the expected transition between bi-polarized state and neutral consensus at the value $n_0^c$, predicted by Eq. \ref{Sc0_transition}. 
Symbols correspond to simulations of the original model, with $N=10000$ agents, and lines represent the analytical solution of the simplified model. 
We can appreciate the full match between simulations and analytical curves. In particular, the simplified model allows a simple interpretation of the dependence of $n_{+/-}(t_f)$ with  $n^i_0$, taking advantage that all agents change their opinion at the same time. 
The bi-polarized state occurs because all neutral agents change their opinion to the dominant extreme opinion ($n_+$ according to Eq. \ref{condicion_inicial}) and therefore: $n_+(t_f)=n_0(0)+n_+(0)=\frac{1+n_0^i}{2}+d$ and $n_-(t_f)=n_-(0)=\frac{1-n_0^i}{2}-d$.

We also look at the temporal evolution of the proportion of agents for the communities, starting from two different initial conditions that correspond to both sides of the transition. 
For $n^i_0=0.3$ (see Fig.\ref{fig4}(b)), we show results for a community that changes opinion at $T_1=t_0\approx 4.17$ reaching a bi-polarized final state. 
For $n^i_0=0.47$ (Fig.\ref{fig4}(c)), there are two communities that change opinions, a first one at $T_1=t_+ \approx3.77$, and a second one at $T_2=\hat{t}_-\approx4.34$, reaching a neutral consensus. 

\begin{figure}[htp]
    \centering
    \includegraphics[width =\columnwidth]{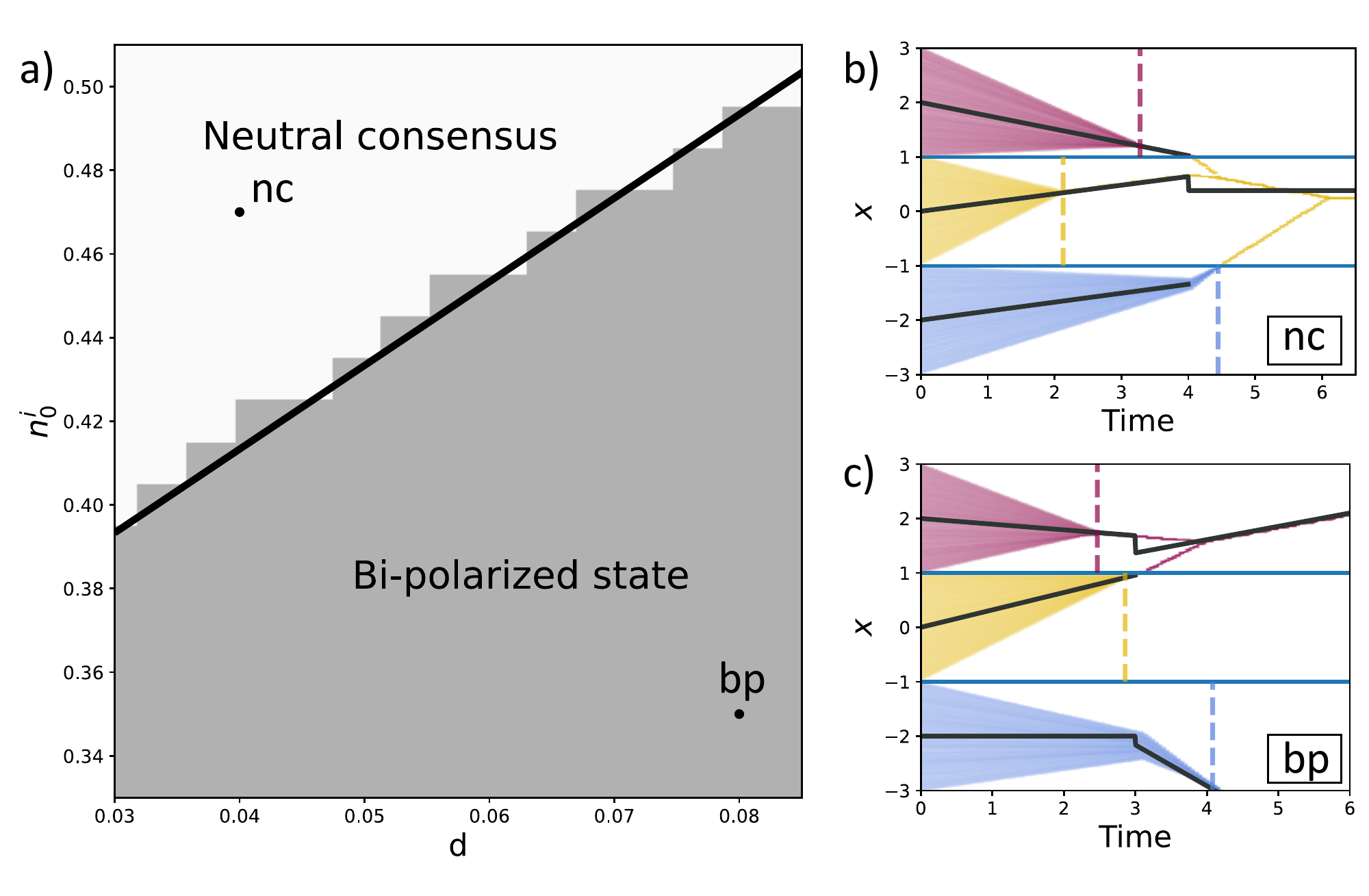}
    \caption{Final state and temporal dynamics. Parameters are set to $L=3$, $C=1$, $r=2$ and $N=10000$. Agents' initial leaning is uniformly distributed for each opinion. 
    Panel a) shows the transition between consensus and bi-polarized final states as a function of $d$ and $n_0^i$. Theoretical prediction for the transition (black line) shows a perfect agreement with simulations. 
    Panels b) and c) show examples of the temporal dynamics for two particular initial conditions: a bi-polarized dynamics ($d=0.04$ and $n_0^i=0.47$), and a consensus dynamics ($d=0.08 $ and $n_0^i=0.35$),  theoretical predictions are in black. 
    }
    \label{fig5}
\end{figure}

Also, we explore the transition between a bi-polarized state and a neutral consensus by varying both parameters: $d$ and $n_0^i$, as shown in Fig. \ref{fig5}(a). We can observe a perfect match between numerical simulations and the analytical predictions of our simplified model, given by  Eq. \ref{Sc0_transition} (black line). 

It should be noticed that our simplified model predicts not only the final state but also the average dynamics of the populations. In Fig. \ref{fig5}(b, c), we show the analytical prediction (black line) and simulations for the agent's leaning distribution (colors), for two particular initial conditions: for  $d=0.04$ and $n_0^i=0.47$ a bi-polarized state is reached (panel (b)), for $d=0.08 $ and $n_0^i=0.35$ the final state is a neutral consensus. 
In Appendix (\ref{appendix:time}), we show that the model's approximation is valid when the concentration time of each community (dashed vertical lines) is smaller than the time for a community to change an opinion. 
This condition is fulfilled when $2rd<n_0^i<\frac{1}{2}$ which indeed happens in the cases demonstrated in Fig. \ref{fig5}(b, c). We observe that when the communities reach a threshold, all agents' leanings have already concentrated in their average, and agents change opinions simultaneously at the same time. Also here, the analytical prediction gives a perfect agreement with simulations.
 
\begin{figure}[htp]
    \centering
    \includegraphics[width =\columnwidth]{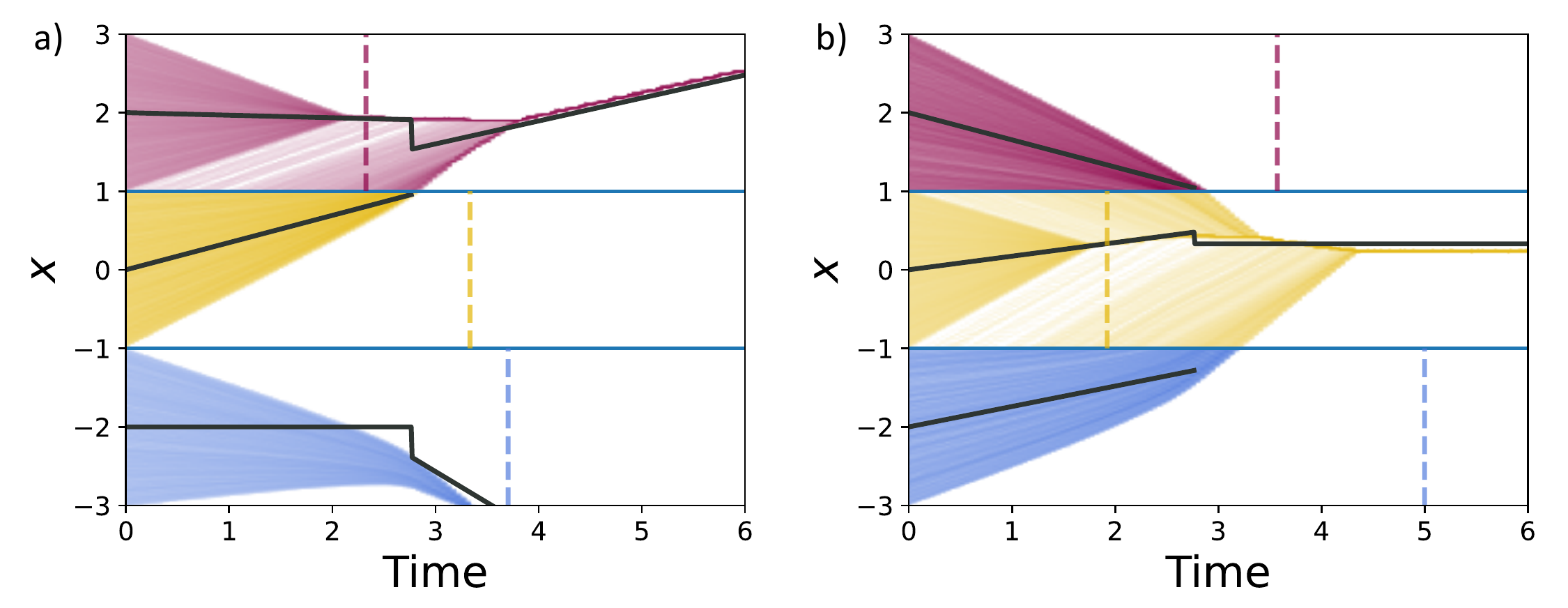}
    \caption{Examples of the temporal dynamics for initial conditions which do not fulfill the condition of convergence given by Eq. \ref{condition}: $d=0.04$ and $n_0^i=0.52$ (panel a), and $d=0.08$ and $n_0^i=0.3$ (panel b). Parameters are set to $L=3$, $C=1$, $r=2$, $N=10000$ and agent's initial leaning for each opinion are uniformly distributed. Although some agents change their opinions before the average leaning reaches the threshold, the approximation still fits very well.}
    \label{fig6}
\end{figure}

Finally, we analyze what happens when the condition provided by Eq. \ref{condition} is not fulfilled and, therefore, the average leaning of one of the communities reaches a threshold before all the leanings have been concentrated. 
In Fig. \ref{fig6}(a), we plot  numerical simulations for initial conditions given by  $d=0.04$ and $n_0^i=0.52$. We can observe that some neutral agents reach positive opinion states before the majority of the community changes their opinion, showing the failures of the simplified model in properly follow the average leaning of each community. 
In panel (b), we plot simulations for initial conditions given by $d=0.08$ and $n_0^i=0.3$. Here we see how some agents with extreme opinions change to neutral before the first average leaning change of opinion. 
However, given that the number of agents following this behavior is small, the proposed approximation still produces a reasonable matching and allows us to predict the final state.

\section{Conclusions}
\label{sec:conclu}

We perform an approximation to resolve the model presented in \cite{BPS} analytically. In this way, we accomplish a drastic reduction of the system's dimension. As a result, we find that the model, on a mesoscopic scale, is governed by very simple rules, which allows us to get an accurate prediction of the dynamics.

We find that, under certain conditions, the dynamics occur on two temporal scales. In a fast one, interactions between agents with the same opinions prevail leading to the formation of groups with the same leaning and the same opinion. Then on a slower scale, interactions between agents with different opinions rule the dynamics, leading the system to the few macroscopic states that fully describe it. Therefore, we can consider only the dynamics of the average leaning for each community as a good approximation for the collective dynamics.

Our main contribution here is to provide an analytic deduction for the average leaning time function from the master equations of the density of the agent's leaning (Eq. \ref{ecuaciones_generales}) and get a simple formula consisting of a piece-wise defined linear function (Eqs. \ref{ecuacion_lineal_1})

The analytical prediction shows significant agreement with the numerical simulations. 
It accurately describes not only the dynamics of the average leaning of each community but also the transition between a neutral consensus state - in which all agent's leanings have the same neutral values- and a bi-polarized state with two communities of agents with the same leaning but contrary opinions. Moreover, we can compute explicitly the values for this transition depending on the initial conditions.

Additionally, we are also able to provide the range of parameters where the separation of the two-time scale assumption applies. Surprisingly, we numerically found that this approximation is good even for parameter values outside this range.

\section{Acknowledgments}

\par This work was supported by Grant PICT-2020-SERIEA-00966 (ANPCyT), Argentina.

\par All authors contributed to the design and implementation of the research, to the analysis of the results and to the writing of the manuscript.

\appendix

\section{Analytical derivation of the master equations}
\label{appendix:equations}

In this appendix, we show that the dynamical equations for the average leaning of agents with the same opinion are governed by Eq. \ref{ecuacion_lineal_1}.

\begin{proof}
 We show the proof only for one case,  the other equations are similar.

To this end, let us compute
\begin{equation}
\frac{d}{dt}( m_+(t))=\frac{1}{n_+(t)} \frac{d}{dt}   \int_C^L x f(x,t) dx.
\label{ecuaciones_centro_masa1}
\end{equation}

As we noted before, if agents of each community have a sufficiently concentrated orientation far from the limit of opinion $C_T$, we can assume that the fraction of agents in each community (in this case $n_{+}(t)$) remains constant. 
In other words, the average leanings are only modified by modifications of agents within the community and not by changes of opinion. 

Let us compute now the right hand side using Eq. \ref{ecuaciones_generales}:
\begin{align*}
 \frac{d}{dt}  \int_C^L x f(x,t) dx=  &
\int_C^L x  \frac{d}{dt} f(x,t)   dx\\
=  &  \int_C^L x
\frac{d}{dx} \Big[   f(x,t) \Big( \int_C^x f(y,t)dy - \int_x^L f(y,t)dy\Big)   \\
&
+  f(x,t) (n_0(t)- n_-(t)) \Big]
dx.
\end{align*}
Integrating by parts, and by using the assumption that the density is concentrated away from the thresholds, and therefore, $f(L,t)=f(C,t)=0$,

\begin{align*}
 \frac{d}{dt}  \int_C^L x f(x,t) dx=
&  - \int_C^L   \Big[ f(x,t) \Big( \int_C^x f(y,t)dy - \int_x^L f(y,t)dy\Big)   +  f(x,t) ( n_0(t) - n_-(t) ) \Big] dx  \\
=& - \int_C^L   \Big[ f(x,t) \Big( \int_C^x f(y,t)dy - \int_x^L f(y,t)dy\Big) \Big]
dx  +  n_+(t) ( n_0(t) - n_-(t) ).
\end{align*}
Therefore, we have
\begin{align*}
 \frac{d}{dt}  m_+(t)=\frac{1}{n_+} \Big(- \int_C^L   \Big[ f(x,t) \Big( \int_C^x f(y,t)dy - \int_x^L f(y,t)dy\Big) \Big]
dx  +  n_+(t) ( n_0(t) - n_-(t) ) \Big).
\end{align*}

Let us show now that the first integral in the right hand side is equal to zero. To this end, let us call
$$
F(x,t)=\int_C^x f(y,t)dy,
$$
and observe that $\frac{d}{dx} F(x,t) = f(x,t)$, and $\int_x^L f(y,t)dy=n_+(t)-F(x,t)$. We rewrite and compute the integral as follows:
 \begin{align*}
  \int_C^L
 f(x,t) \Big( 2F(x,t) - n_+(t) \Big)  dx
= &
\frac{1}{4}  \int_C^L \frac{d}{dx}
\Big( 2F(x,t)  - n_+(t) \Big)^2  dx \\
=&
\frac{1}{4}\Big( 2F(x,t) - n_+(t) \Big)^2\Big|_L^C
\\ =  &
\frac{1}{4} \Big( 2F(L,t) - n_+(t) \Big)^2 -
\frac{1}{4} \Big( 2F(C,t) - n_+(t) \Big)^2 \\
= & 0,
\end{align*}
since $F(L,t)=n_+(t)$ and $F(C,t)=0$.

Therefore, 
\begin{align*}
 \frac{d}{dt}  m_+(t)=-\frac{1}{n_+}(  n_+(t) ( n_0(t) - n_-(t) ) )= n_-(t)-n_0(t)   
\end{align*}

and the proof is finished.
\end{proof}

\section{Conditions for the validity of the simplified model}
\label{appendix:time}

An important hypothesis for the approximation is that communities are concentrated rapidly compared with the move of the average leaning, and therefore, we can assume the entire community is concentrated near the average leaning, and agents change opinions altogether at the same time. 
This appendix will will give the necessary conditions to fulfill this hypothesis.
We consider the results of \cite{PSB} where the authors analyze a similar model with just one community. For only one community and constant opinion’s initial distribution, they show that the solution blows up when $t$ reaches 1.

We use this result for the density in Eq. \ref{ecuaciones_generales}.
Considering that the only interactions that tend to concentrate a community are the interactions inter communities, the agents concentrate at a different time for each community. 
This time will be $\tilde{t_i}=\frac{1}{n_i}$ for $i \in \{-,+,0\}$. From that time on, communities are concentrated near the center of mass and moved together.
For the assumption made in the previous section to be fulfilled, we need the convergence time of each community to occur before the change of opinion. So, when the mass center reaches the threshold, the community is concentrated. therefore, $\tilde{t}_i$ needs to be less than its respective time of change of opinion $t_i$ given in Eq. \ref{tiempo}. For example, for $i=+$ we have

\begin{equation*}
\begin{split}
t_+&>\tilde{t}_+\\
\frac{C-m_+(0)}{n_{-}(0)-n_{0}(0)}&>\frac{ 1}{n_{+}(0)},\\
\frac{1}{n_{-}(0)-n_{0}(0)}&>\frac{1}{n_{+}(0)},\\
n_{+}(0)&>n_{-}(0)-n_{0}(0) ,
\label{times}
\end{split}
\end{equation*}
where we use that $C-m_{+}(0)=1$ for the constant opinion initial distribution and we assume $C=2$ and $L=3$. Equivalently, for the other communities, we obtain

\begin{align*}
n_{+}(0)&>n_{-}(0)-n_{0}(0),\\
n_{-}(0)&>n_{0}(0)-n_{+}(0),\\
n_{0}(0)&>r(n_{+}(0)-n_{-}(0)).
\end{align*}

If we replace these conditions with the initial free parameters in \ref{condicion_inicial}, we get that the condition that must be satisfied in order to be the hypotheses of the previous results is the following
\begin{equation}
\label{condition}
2rd<n^i_0<\frac{1}{2},
\end{equation}
and we get upper bounds for the initial distributions, and the parameters $r$ and $d$.


\begin{thebibliography}{10}

\bibitem{Ianellietal2021}
L. Iannelli, B. Biagi, and M. Meleddu.
\newblock Public opinion polarization on immigration in italy: the role of
  traditional and digital news media practices.
\newblock {\em The Communication Review}, 24(3):244--274, 2021.

\bibitem{Aaron2022}
A. McCright and R. Dunlap.
\newblock The politicization of climate change and polarization in the american
  public's views of global warming, 2001–2010.
\newblock {\em The Sociological Quarterly}, 52(2):155--194, 2011.

\bibitem{Milligan2020}
S. Milligan.
\newblock The political divide over the coronavirus. {US} news \& {W}orld
  report.
\newblock
  \url{https://www.usnews.com/news/politics/articles/2020-03-18/the-political-divide-over-the-coronavirus},
  2020.

\bibitem{Roberts2020}
D. Roberts.
\newblock Partisanship is the strongest predictor of coronavirus response.
\newblock
  \url{https://www.vox.com/science-and-health/2020/3/31/21199271/coronavirus-in-us-trump-republicans-democrats-survey-epistemic-crisis},
  2020.

\bibitem{flache2017}
A. Flache, M. M\"{a}s, T. Feliciani, E. Chattoe-Brown,
  G. Deffuant, S. Huet, and J. Lorenz.
\newblock Models of social influence: Towards the next frontiers.
\newblock {\em Journal of Artificial Societies and Social Simulation}, 20(4):2,
  2017.

\bibitem{kozitsin2022formal}
I. Kozitsin.
\newblock Formal models of opinion formation and their application to real
  data: evidence from online social networks.
\newblock {\em The Journal of Mathematical Sociology}, 46(2):120--147, 2022.

\bibitem{Schelling}
T. Schelling.
\newblock {\em Micromotives and macrobehavior}.
\newblock WW Norton \& Company, 1978.

\bibitem{Granovetter}
M. Granovetter.
\newblock Threshold models of collective behavior.
\newblock {\em American Journal of Sociology}, 83(6):1420--1443, 1978.

\bibitem{Vickers}
D. Vickers and M. Lee.
\newblock Dynamic models of simple judgments: I. properties of a
  self-regulating accumulator module.
\newblock {\em Nonlinear Dynamics, Psychology, and Life Sciences},
  2(3):169--194, 1998.

\bibitem{Smith}
P. Smith and R. Ratcliff.
\newblock Psychology and neurobiology of simple decisions.
\newblock {\em Trends in Neurosciences}, 27(3):161--168, 2004.

\bibitem{Cinelli}
M. Cinelli, G. De~Francisci Morales, A. Galeazzi, W.
  Quattrociocchi, and M. Starnini.
\newblock The echo chamber effect on social media.
\newblock {\em Proceedings of the National Academy of Sciences},
  118(9):e2023301118, 2021.

\bibitem{Abelson}
R. Abelson.
\newblock Mathematical models in social psychology.
\newblock volume~3 of {\em Advances in Experimental Social Psychology}, pages
  1--54. Academic Press, 1967.

\bibitem{DeGroot}
M. DeGroot.
\newblock Reaching a consensus.
\newblock {\em Journal of the American Statistical Association},
  69(345):118--121, 1974.

\bibitem{Weisbuch2}
G.~Weisbuch, G.~Deffuant, F.~Amblard, and J.-P. Nadal.
\newblock Interacting agents and continuous opinions dynamics.
\newblock In Robin Cowan and Nicolas Jonard, editors, {\em Heterogenous Agents,
  Interactions and Economic Performance}, pages 225--242, Berlin, Heidelberg,
  2003. Springer Berlin Heidelberg.

\bibitem{Burnstein}
E. Burnstein and A. Vinokur.
\newblock Persuasive argumentation and social comparison as determinants of
  attitude polarization.
\newblock {\em Journal of Experimental Social Psychology}, 13(4):315--332,
  1977.

\bibitem{Sunstein}
C. Sunstein.
\newblock The law of group polarization.
\newblock {\em John M. Olin Program in Law and Economics Working Paper No.91},
  (91), 1999.

\bibitem{Sampedro}
V. Sampedro and F.S. Pérez.
\newblock The 2008 spanish general elections: “antagonistic bipolarization”
  geared by presidential debates, partisanship, and media interests.
\newblock {\em The International Journal of Press/Politics}, 13(3):336--344,
  2008.

\bibitem{Dandekar}
P. Dandekar, A. Goel, and D.T. Lee.
\newblock Biased assimilation, homophily, and the dynamics of polarization.
\newblock {\em Proceedings of the National Academy of Sciences},
  110(15):5791--5796, 2013.

\bibitem{Krueger}
T. Krueger, J. Szwabiński, and T. Weron.
\newblock Conformity, anticonformity and polarization of opinions: Insights
  from a mathematical model of opinion dynamics.
\newblock {\em Entropy}, 19(7), 2017.

\bibitem{Jager}
W. Jager and F. Amblard.
\newblock Uniformity, bipolarization and pluriformity captured as generic
  stylized behavior with an agent-based simulation model of attitude change.
\newblock {\em Computational \& Mathematical Organization Theory},
  10(4):295--303, 2005.

\bibitem{ben1996coarsening}
E. Ben-Naim, L. Frachebourg, and P.L. Krapivsky.
\newblock Coarsening and persistence in the voter model.
\newblock {\em Physical Review E}, 53(4):3078, 1996.

\bibitem{Clifford}
P. Clifford and A. Sudbury.
\newblock A model for spatial conflict.
\newblock {\em Biometrika}, 60(3):581--588, 1973.

\bibitem{Holley}
R.A. Holley and T.M. Liggett.
\newblock Ergodic theorems for weakly interacting infinite systems and the
  voter model.
\newblock {\em The Annals of Probability}, 3(4):643--663, 1975.

\bibitem{Cox}
J.T. Cox and D. Griffeath.
\newblock Diffusive clustering in the two dimensional voter model.
\newblock {\em The Annals of Probability}, 14(2):347--370, 1986.

\bibitem{Liggett}
T.M. Liggett.
\newblock {\em Interacting Particle Systems}, volume 276.
\newblock Springer Science \& Business Media, 2012.

\bibitem{Sire}
C. Sire and S.N. Majumdar.
\newblock Coarsening in the q-state {P}otts model and the {I}sing model with
  globally conserved magnetization.
\newblock {\em Physical Review E}, 52(1):244, 1995.

\bibitem{sznajd2000opinion}
K. Sznajd-Weron and J. Sznajd.
\newblock Opinion evolution in closed community.
\newblock {\em International Journal of Modern Physics C}, 11(06):1157--1165,
  2000.

\bibitem{deffuant2000mixing}
G. Deffuant, D. Neau, F. Amblard, and G. Weisbuch.
\newblock Mixing beliefs among interacting agents.
\newblock {\em Advances in Complex Systems}, 3(01n04):87--98, 2000.

\bibitem{weisbuch2004bounded}
G. Weisbuch.
\newblock Bounded confidence and social networks.
\newblock {\em The European Physical Journal B}, 38(2):339--343, 2004.

\bibitem{lorenz2007continuous}
J. Lorenz.
\newblock Continuous opinion dynamics under bounded confidence: A survey.
\newblock {\em International Journal of Modern Physics C}, 18(12):1819--1838,
  2007.

\bibitem{hegselmann2002opinion}
R. Hegselmann, U. Krause, et~al.
\newblock Opinion dynamics and bounded confidence models, analysis, and
  simulation.
\newblock {\em Journal of artificial societies and social simulation}, 5(3),
  2002.

\bibitem{amblarddeffuant2004}
F.~Amblard and G.~Deffuant.
\newblock The role of network topology on extremism propagation with the
  relative agreement opinion dynamics.
\newblock {\em Physica A: Statistical Mechanics and its Applications},
  343:725--738, 2004.

\bibitem{deffuant2006}
G. Deffuant.
\newblock Comparing extremism propagation patterns in continuous opinion
  models.
\newblock {\em Journal of Artificial Societies and Social Simulation}, 9(3):8,
  2006.

\bibitem{Bellomo}
N. Bellomo, G.~A. Marsan, and A. Tosin.
\newblock {\em Complex systems and society: Modeling and simulation}, volume~2.
\newblock Springer, 2013.

\bibitem{Pareschi}
L. Pareschi and G. Toscani.
\newblock {\em Interacting multiagent systems: {K}inetic equations and {M}onte
  Carlo methods}.
\newblock OUP Oxford, 2013.

\bibitem{Toscani}
G. Toscani et~al.
\newblock Kinetic models of opinion formation.
\newblock {\em Communications in Mathematical Sciences}, 4(3):481--496, 2006.

\bibitem{BPS}
P. Balenzuela, J.~P. Pinasco, and V. Semeshenko.
\newblock The undecided have the key: Interaction-driven opinion dynamics in a
  three state model.
\newblock {\em PLOS ONE}, 10(10):1--21, 10 2015.

\bibitem{BSNB}
F. Barrera~Lemarchand, V. Semeshenko, J. Navajas, and
  P. Balenzuela.
\newblock Polarizing crowds: Consensus and bipolarization in a persuasive
  arguments model.
\newblock {\em Chaos: An Interdisciplinary Journal of Nonlinear Science},
  30(6):063141, 2020.

\bibitem{deLaLama}
M.~de~La~Lama, I.~Szendro, J.~Iglesias, and H.~Wio.
\newblock {V}an {K}ampen's expansion approach in an opinion formation model.
\newblock {\em The European Physical Journal B-Condensed Matter and Complex
  Systems}, 51(3):435--442, 2006.

\bibitem{Couzin}
I.~D. Couzin, C.~C. Ioannou, G. Demirel, T. Gross, C.~J.
  Torney, A. Hartnett, L. Conradt, S.~A. Levin, and N.~E.
  Leonard.
\newblock Uninformed individuals promote democratic consensus in animal groups.
\newblock {\em Science}, 334(6062):1578--1580, 2011.

\bibitem{Sobkowicz}
P. Sobkowicz.
\newblock Discrete model of opinion changes using knowledge and emotions as
  control variables.
\newblock {\em PLOS ONE}, 7(9):1--16, 09 2012.

\bibitem{Vazquez}
F.~Vazquez and S.~Redner.
\newblock Ultimate fate of constrained voters.
\newblock {\em Journal of Physics A: Mathematical and General},
  37(35):8479--8494, aug 2004.

\bibitem{Svenkeson}
A.~Svenkeson and A.~Swami.
\newblock Reaching consensus by allowing moments of indecision.
\newblock {\em Scientific Reports}, 5(1):1--9, 2015.

\bibitem{Singh}
P.~Singh, S.~Sreenivasan, B.~K. Szymanski, and G.~Korniss.
\newblock Competing effects of social balance and influence.
\newblock {\em Phys. Rev. E}, 93:042306, Apr 2016.

\bibitem{Marvel}
S.~A. Marvel, H. Hong, A. Papush, and S.~H. Strogatz.
\newblock Encouraging moderation: Clues from a simple model of ideological
  conflict.
\newblock {\em Phys. Rev. Lett.}, 109:118702, Sep 2012.

\bibitem{Galam1}
S. Galam.
\newblock The drastic outcomes from voting alliances in three-party democratic
  voting (1990→ 2013).
\newblock {\em Journal of Statistical Physics}, 151(1):46--68, 2013.

\bibitem{Galam2}
S. Gekle, L. Peliti, and S. Galam.
\newblock Opinion dynamics in a three-choice system.
\newblock {\em The European Physical Journal B-Condensed Matter and Complex
  Systems}, 45(4):569--575, 2005.

\bibitem{Galam3}
S. Galam.
\newblock Sociophysics: A review of galam models.
\newblock {\em International Journal of Modern Physics C}, 19(03):409--440,
  2008.

\bibitem{Galam4}
S. Galam.
\newblock {\em Sociophysics: A Physicist's Modeling of Psycho-political
  Phenomena}.
\newblock Springer, 2016.

\bibitem{PSB}
J.P. Pinasco, V. Semeshenko, and P. Balenzuela.
\newblock Modelling opinion dynamics: Theoretical analysis and continuous
  approximation.
\newblock {\em Chaos, Solitons \& Fractals}, 98:210--215, 05 2017.

\end{thebibliography}

\end{document}